\documentclass[12pt]{iopart}
\usepackage{hyperref}
\usepackage{graphicx}
\usepackage[caption=false]{subfig}

\begin{document}
\title[Annihilation dynamics of stringlike topological defects in a nematic LLC]{Annihilation dynamics of stringlike topological defects in a nematic lyotropic liquid crystal}

\author{R. R. Guimar\~aes, R. S. Mendes, P. R. G. Fernandes and H. Mukai}

\address{Departamento de F\'isica, Universidade Estadual de Maring\'a, Av. Colombo - 5790, 87020-900, Maring\'a, PR, Brazil}

\ead{pricardo@dfi.uem.br}

\begin{abstract}
Topological defects can appear whenever there is some type of ordering. Its ubiquity in nature has been the subject of several studies, from early Universe to condensed matter. In this work, we investigated the annihilation dynamics of defects and antidefects in a lyotropic nematic liquid crystal (ternary mixture of potassium laurate, decanol and deionized-destilated water) using the polarized optical light microscopy technique. We analyzed \textit{Schlieren} textures with topological defects produced due to a symmetry breaking in the transition of the isotropic to nematic calamitic phase after a temperature quench. As result, we obtained for the distance $D$ between two annihilating defects (defect-antidefect pair), as a function of time $t$ remaining for the annihilation, the scaling law $D \propto t^{\alpha}$, with $\alpha = 0.390$ and standard deviation $\sigma = 0.085$. Our findings go in the direction to extend experimental results related to dynamics of defects in liquid crystals since only thermotropic and polymerics ones had been investigated. In addition, our results are in good quantitative agreement with previous investigations on the subject.
\end{abstract}

\maketitle

\section{Introduction}
Liquid crystals \cite{degennes, chandra, lavrentovich} are formed by anisotropic molecular units and share properties of isotropic liquids and ordered crystals as flow and molecular ordering, respectively. They can be classified, according to its basic constituents, as thermotropics (organic molecules), lyotropics (micelles formed by amphiphilic molecules) \cite{fneto}, followed by the polymerics (polymers) \cite{march} and metallotropics (composed of both organic and inorganic molecules) \cite{martin}. The phase transitions in these materials occur by changes in temperature, concentration, or pressure, and various mesophases can be observed, generally identified by different characteristic optical textures \cite{dierking}. The most known liquid-crystalline mesophase is the nematic one, which presents orientational order of its constituents and may have one (uniaxial) or two (biaxial) optical axes. The uniaxial nematic mesophase is further classified in \textit{nematic calamitic}, when the molecules/micelles are rod shaped, and in \textit{nematic discotic}, when the molecules/micelles have the shape of a disc. One of the ways to characterize this mesophase is via the birefringence ($\Delta n$), for instance, when $\Delta n \neq 0$, the liquid-crystalline material shows the nematic mesophase and, if $\Delta n =0$, it shows the isotropic phase. Experimentally, the nematic mesophase can be obtained by decreasing the temperature from the isotropic one.

The ordering process of systems after a phase transition have been largely studied in condensed matter physics \cite{chuang, turok, pargellis, finn, yurke, toyoki, zapotocky, marshall, dutta, oliveira}, through theories and experiments. Some of the main aspects of this process is the behaviour of the defects present in the system such as its formation and annihilation. Topological defects can appear whenever there is some sort of ordering. Its ubiquity in nature has been the subject of several studies \cite{kibble, mermin, charlier, vakaryuk, figueiras, petit, abu, carvalho} in addition to those ones in condensed matter. Here we are interested in topological defects in a more restritive context, one related to liquid crystals \cite{shiwaku, blundell, cypt, ding, zapotocky2, mendez, minoura, minoura2, wang, rojas, denniston, svetec}. In this scenario, universal behaviours can be identified, for instance, the similarity with defects formed in the early Universe has provided studies using a liquid crystal as a cosmological laboratory \cite{chuang, digal, mukai, pawel, dhara}. In the liquid-crystalline systems, generally, the defects arise after an isotropic-nematic phase transition, i.e., a symmetry breaking of isotropic phase \cite{degennes, chandra, lavrentovich}. The defects in a liquid crystal basically are regions where the director $\vec{n}$ is not defined. In general, $\vec{n}$ represents the average molecular alignment and shows an inversion symmetry ($\vec{n} \leftrightarrow -\vec{n}$). Due to the birefringence of the liquid-crystalline mesophases, and their extremely small elastic constants, they are ideal materials to study the defects dynamics experimentally, because the defect's motion can be followed in real-time through polarized optical light microscopy technique on practical spatial and temporal scales.

Much of the reported experimental work about the annihilation of defects \cite{chuang, turok, pargellis, finn, yurke, marshall, cypt, mendez, minoura, wang, blanc, nagaya} was made using thermotropic liquid crystals, where stringlike defects, pointlike defects, and defects loop were analyzed. In thermotropic polyester materials, the annihilation of defects have been observed too \cite{shiwaku, ding, wang}. The annihilation of defects has also been analyzed in numeric simulations \cite{yurke, zapotocky, oliveira, svetec, svensek}. However, an experimental investigation of this matter on lyotropic liquid crystals is an open question yet. In the direction to fill this gap, we used a lyotropic liquid crystal (KL-DeOH-deionized/destilated H$_2$O) at nematic calamitic phase to analyze the annihilation dynamics of stringlike defects. We produced the defects by a temperature quench, and focused our attention in the evolution of the distance between defects in process of annihilation. In order to accomplish our goal, the polarized light optical microscopy technique was used. We divided our work as follows: Basic theoretical aspects are showed in section \ref{theory}; the description of the experiment is in section \ref{exp}; and the results and discussion are presented in section \ref{resdis}. The last section (section \ref{conc}) is devoted to the conclusions.

\section{Theory}\label{theory}

\begin{figure}
\center
\subfloat[]{\label{thermo}\includegraphics[width=6.5cm]{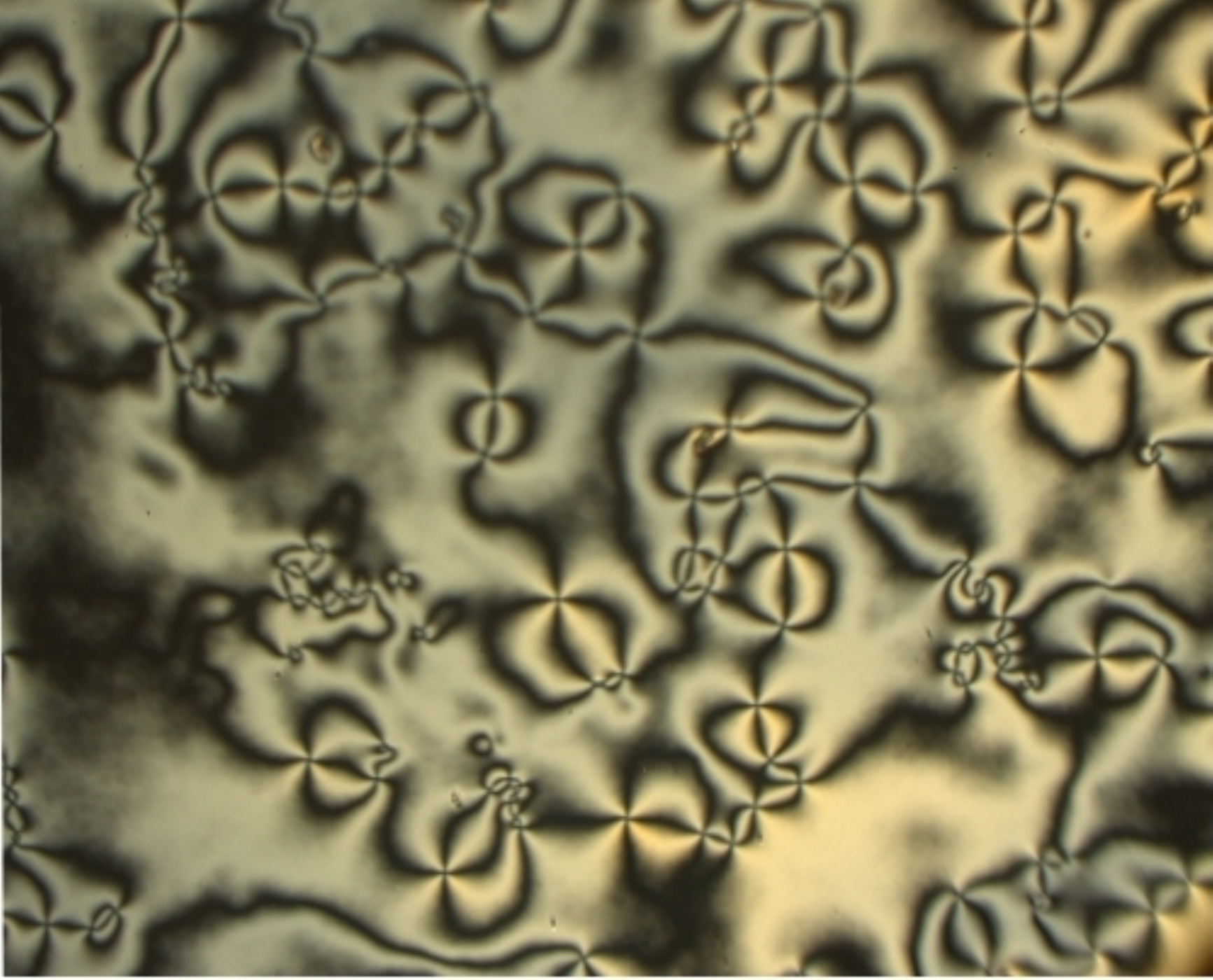}}
\qquad
\subfloat[]{\label{lyo}\includegraphics[width=6.5cm]{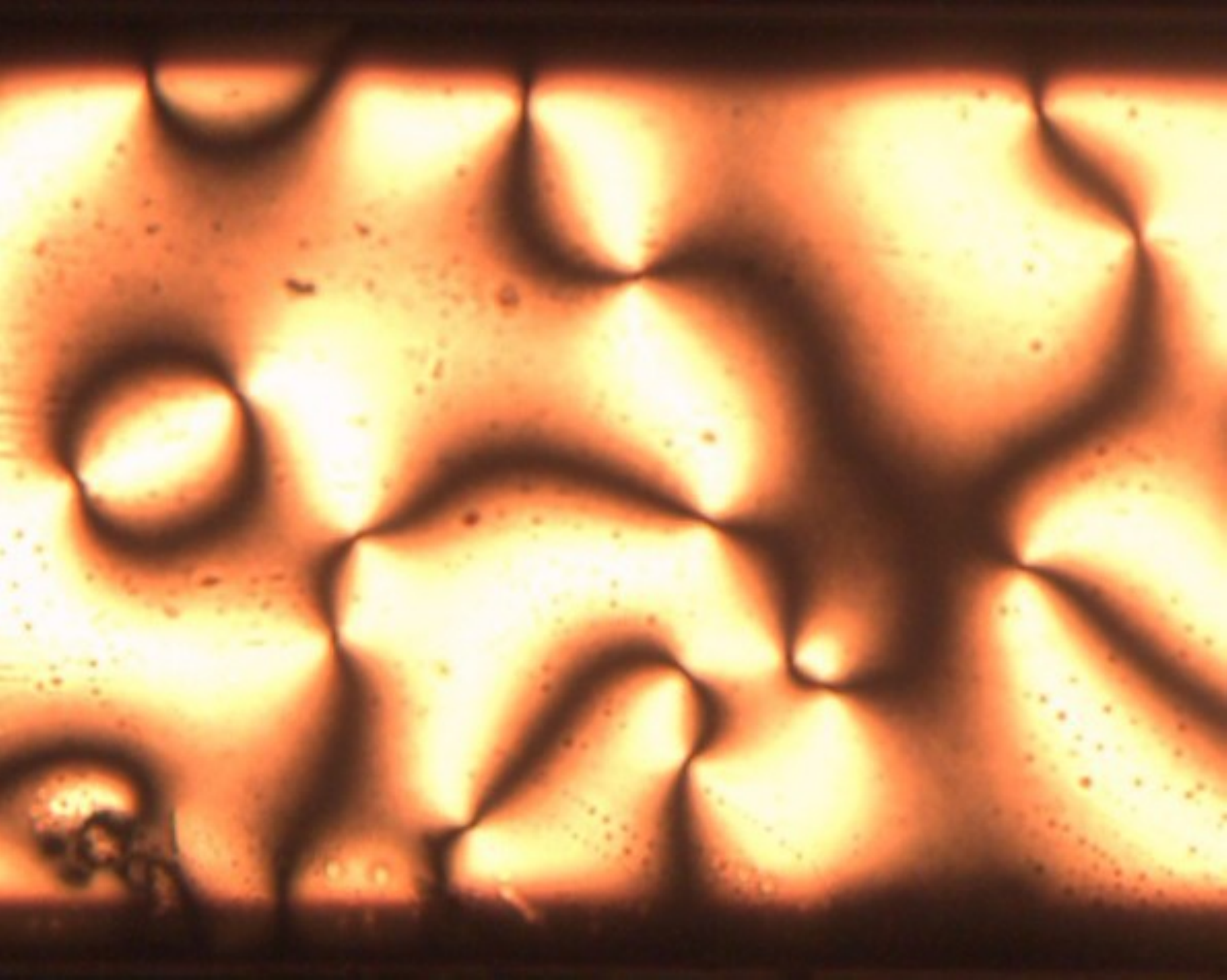}}
\caption{(a) \textit{Schlieren} texture of a thermotropic liquid crystal after the isotropic-nematic transition \cite{gsdias}. (b) \textit{Schlieren} texture of a lyotropic ternary mixture \cite{renato}.}
\label{schlieren}
\end{figure}

One typical liquid-crystalline texture, called \textit{Schlieren} texture, obtained by the polarized optical light microscopy technique, displays brushes that converge at singular points \cite{dierking}, in a bi-dimensional configuration (figure \ref{schlieren}). These singular points represent topological defects of different types and signs \cite{degennes, chandra, lavrentovich}, which are characterized according to the arrangement of $\vec{n}$ in the vicinity of the singularities. The director configuration around a defect can be studied from the elastic energy density, called Frank's energy density \cite{degennes, chandra, lavrentovich, barbero}, 
\begin{eqnarray} \label{eq1}
\mathcal{F}_V &=& \frac{1}{2} K_{11} [\nabla \cdot \vec{n}]^2 + \frac{1}{2} K_{22}[\vec{n} \cdot (\nabla \times \vec{n})]^2 + \frac{1}{2} K_{33} [\vec{n} \times (\nabla \times \vec{n})]^2,
\end{eqnarray} 
where $K_{11}$, $K_{22}$, and $K_{33}$ represent the splay, twist, and bend elastic constants respectively. These constants, which are positives, have dimension of energy per length and are temperature dependent. Its typical values are about 10$^{-6}$ \textit{dynas} for both thermotropic \cite{degennes, kleman} and lyotropic \cite{kroin} liquid crystals. 

\begin{figure}[h]
\center
\subfloat[$s = + \frac{1}{2}$]{\label{defect-a}\includegraphics[scale=0.45]{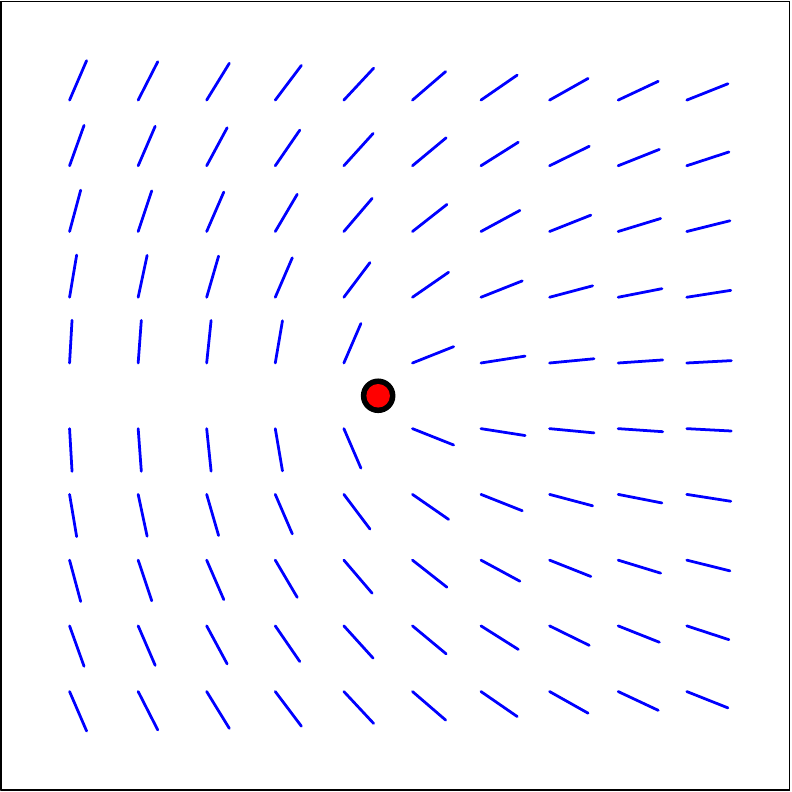}}
\hspace{0.35cm}
\subfloat[$s = -\frac{1}{2}$]{\label{defect-b}\includegraphics[scale=0.45]{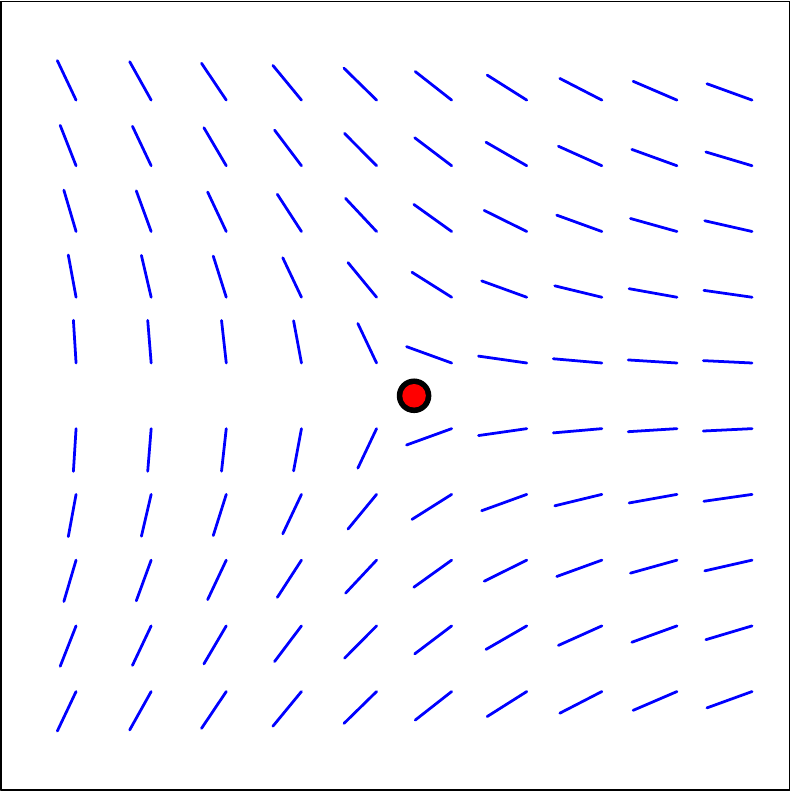}}
\\
\subfloat[$s = +1$]{\label{defect-c}\includegraphics[scale=0.45]{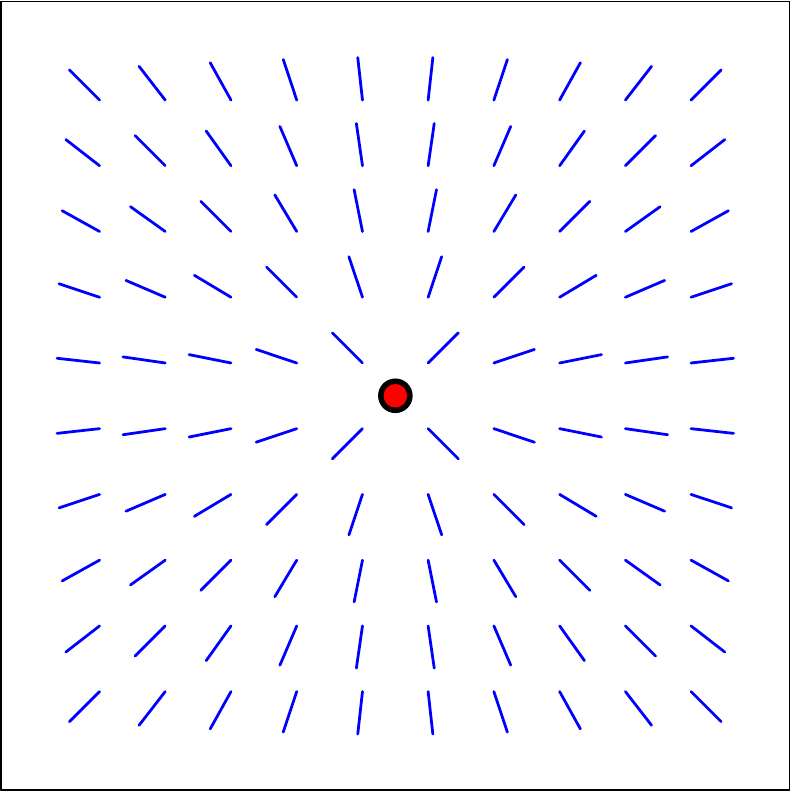}}
\hspace{0.35cm}
\subfloat[$s = -1$]{\label{defect-d}\includegraphics[scale=0.45]{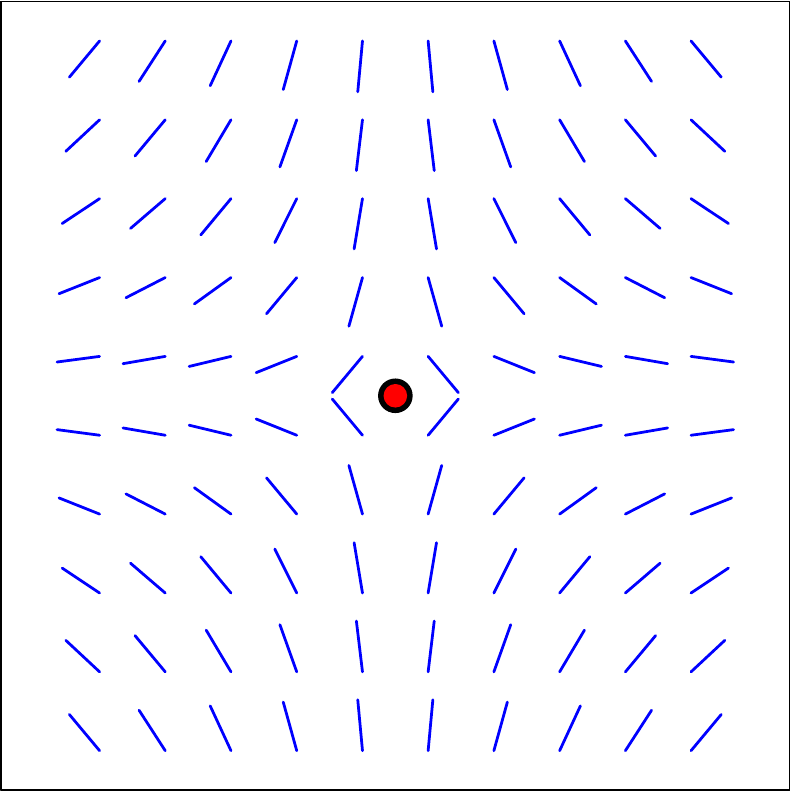}}
\caption{Sketch of the director configurations around a defect for the case $\theta_0 = 0$ \cite{degennes, chandra, lavrentovich}.}
\label{sketch}
\end{figure}

Considering $\vec{n}$ in two dimensions, the solution that minimizes the free energy described in equation (\ref{eq1}), usually obtained by assuming the one-constant approximation for the elastic constants ($K_{11} = K_{22} = K_{33} = K$), is given by \begin{eqnarray} \label{eq2}
\theta(x,y) = s \phi + \theta_0.
\end{eqnarray}
This solution is correct from the point of view of symmetries and assumes the defect at the origin of $x$-$y$ plane, with the $z$-axis perpendicular. $\theta(x,y)$ is the angle between $\vec{n}$ and $x$ axis, and $\phi$ is given by $\tan^{-1} y/x$. $\theta_0$ changes from 0 to $\pi$ \cite{degennes, chandra, lavrentovich}. The $\vec{n}$ orientation changes by $2 \pi s$ on going round the singularity. The $s$ parameter represents the defect's type (topological charge or strength of the defect), assuming values of $\pm 1/2, \pm 1, \pm 3/2, \ldots$, but only defects of strength $s = +1/2, -1/2, +1$ and $-1$ are generally observed \cite{chandra}. Usually, defects with $s = \pm 1/2$ ($s = \pm 1$) are referred as stringlike (pointlike) defects. By convention, if $s$ is positive (negative), the singularity is a \textit{defect} (an \textit{antidefect}). Figure \ref{sketch} shows a sketch of the possible director configurations around a defect. Note that, at the center of the defect, there is a region known as \textit{defect's core} with size of approximately $10^{-8}m$ \cite{chandra}. In general, in a thermotropic liquid crystal, the defects are mostly pointlike (figure \ref{thermo}) and identified by four brushes meeting in a point of the \textit{Schlieren} textures \cite{degennes, chandra, lavrentovich, dierking}. In the lyotropics, stringlike defects (figure \ref{lyo}) are more common and are identified by two brushes meeting in a point of the \textit{Schlieren} textures \cite{degennes, chandra, lavrentovich}. As one can see from figure \ref{schlieren}, both types of defects are connected via the brushes. 

Note also that the sum over all defect's strengths $s$ of a sample tends to zero, leading to a law of conservation of the topological charge \cite{lavrentovich}. The defects distribution is dynamic. In fact, defects of equal strength and opposite sign attract each other going toward mutual annihilation.

Theoretically, as indicated in the \cite{zapotocky}, by using dimensional analysis for two dimensional systems with topological defects, Lifschitz indicated that the scaling law for annihilating defects goes as or slower than $t^{1/2}$, where $t$ is the time remaining for the annihilation. 
More recently, it was obtained equations for the separation $D$ between a defect and an antidefect as a function of $t$. This was made starting from the equation of motion for an isolated defect-antidefect pair, which follows from equating the attractive and frictional forces acting on each defect. If the elastic attractive force is taken to be \cite{degennes, finn, yurke} $F_{\rm at} \propto -1/D$, and the frictional one is taken to be \cite{pargellis} $F_{\rm fr} \propto v $, where $v=(1/2) (\rmd D/\rmd t)$ is the defect's velocity, one obtains $D(t) \propto t^{1/2}$ (see, for instance, Pargellis and co-authors \cite{pargellis}). Other behaviours for $D(t)$ slower than $t^{1/2}$ can be obtained considering corrections to the above forces (see, for instance, Yurke and co-authors \cite{yurke}).

As we shall see in the discussion, numerical simulations and experimental results for liquid crystals \cite{turok, pargellis, finn, yurke, zapotocky, marshall, dutta, oliveira, shiwaku, blundell, cypt, ding, zapotocky2, mendez, minoura, minoura2, wang, rojas, denniston, svetec} support that the annihilation defect-antidefect is well adjusted by the scaling law \begin{eqnarray}
\label{eq4}
D(t) \propto t^{\alpha},
\end{eqnarray}
where $\alpha$ is a constant and $t$ is time remaining for the annihilation. Here, we intend to obtain experimentally the scaling exponent in equation (\ref{eq4}) for the annihilation of such pairs in the lyotropic liquid crystal specified in section \ref{exp}.

\section{Experiment}\label{exp}

The liquid crystal used in our analysis was the ternary lyotropic mixture formed by potassium laurate (KL), decanol (DeOH) and deionized/destiled water (H$_2$O), with the following concentrations in weight percent: KL $\rightarrow$ 27.49\%, DeOH $\rightarrow$ 6.24\% and H$_2$O $\rightarrow$ 66.27\%. 

When an abrupt change of the free energy of the system occurs, defects are usually formed. This abrupt change can be induced, for instance, by pressure \cite{turok, pargellis}, temperature quenches \cite{pargellis, wang, mukai}, and external fields \cite{marshall, oliveira, dhara}. In this work, we investigated the annihilation of defect-antidefect pairs formed due to a temperature quench. Here, only the bulk of the sample was investigated, i.e., the defects analyzed had no influence of the anchoring energy \cite{evangelista}.

The technique employed to observe the defects was the polarized optical light microscopy \cite{lavrentovich, mukai, dhara}. The sample was placed in a glass capillary with 100$\mu$m light path and, to avoid changes in the mixture concentration, its borders were sealed with a paraffin film followed by a nail polish coat. Then, the sample was analyzed between crossed polarizers in an optical microscope Leica DM2500P (5$\times$ objective and 10$\times$ ocular) connected to a charge coupled device (CCD) camera DFC290. The laboratory frame axes were defined as follows: $x$ is the long axis of the glass capillary and $z$ is the axis normal to its largest surface. The heating of the sample was controlled by using a hotstage with a X-Y micropositioner (INSTEC-HCS302, $10^{-3}\,^{\circ}\mathrm{C}$ of precision) connected to a computer and, for the cooling, we used a water bath (accuracy of $10^{-3}\,^{\circ}\mathrm{C}$).

The glass capillary was put in the hotstage around $25\,^{\circ}\mathrm{C}$. At this temperature, the sample presents a nematic calamitic texture. Thereafter, the temperature was increased up to $50\,^{\circ}\mathrm{C}$ in order to reach the isotropic phase. To obtain the desired phase transition (around $40\,^{\circ}\mathrm{C}$), we made a quench to return to a temperature around $25\,^{\circ}\mathrm{C}$. After about 5 hours, the topological defects became pronounced, that is, they were well visualized with CCD camera. Figure \ref{lyo} shows a typical texture of the lyotropic mixture presenting the defects. The dark brushes in figure \ref{lyo} are regions where the mean micellar alignment is parallel or perpendicular to the plane of polarization of the incident light.

The procedure to analyze the annihilation dynamics was to take a sequence of photos in a specific region of the glass capillary until the maximum of annihilations of defect-antidefect pairs were carried out. The data acquisition was focused on the position of the annihilating defects in the picture. The time intervals of 60, 90, 120, 150, 240, 300 and 600 seconds between each photograph were used. We considered annihilations ocurred from 5340 s to 119 700 s remaining for the end of the process. The resolution of the photographs was 1024 $\times$ 768 pixels.

\section{Results and Discussion}\label{resdis}

\begin{figure}[h]
\center
\subfloat[t=17910 s]{\label{17910}\includegraphics[scale=0.45]{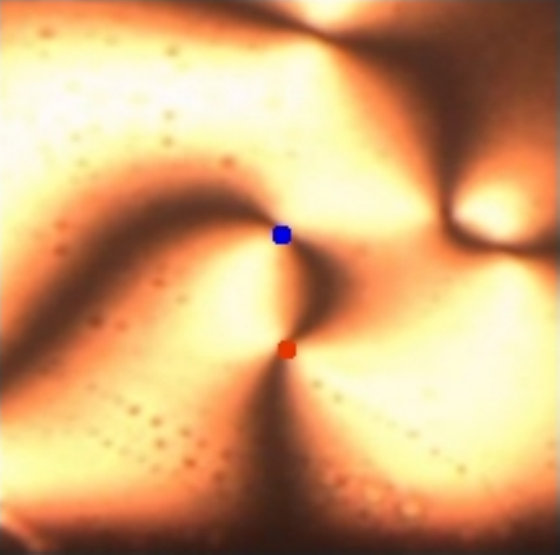}}
\hspace{0.35cm}
\subfloat[t=14400 s]{\label{14400}\includegraphics[scale=0.45]{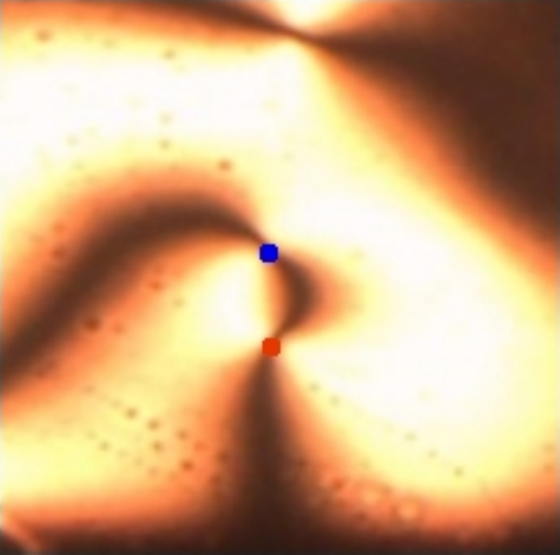}}
\hspace{0.35cm}
\subfloat[t=10800 s]{\label{10800}\includegraphics[scale=0.45]{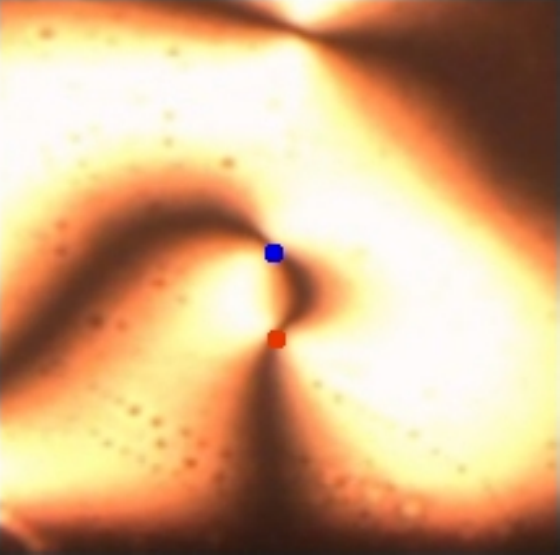}}
\\
\subfloat[t=7200 s]{\label{7200}\includegraphics[scale=0.45]{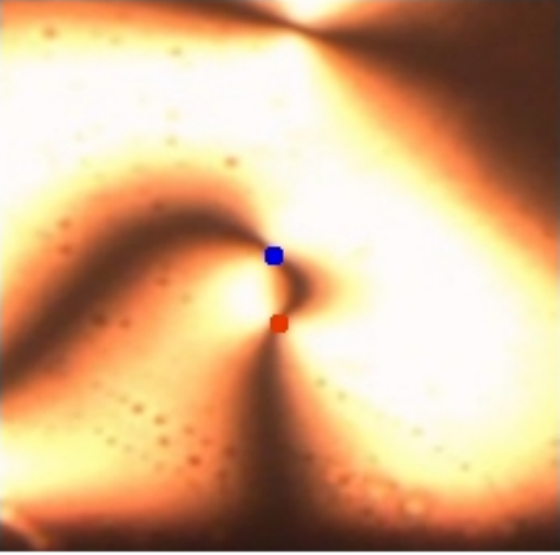}}
\hspace{0.35cm}
\subfloat[t=1800 s]{\label{1800}\includegraphics[scale=0.45]{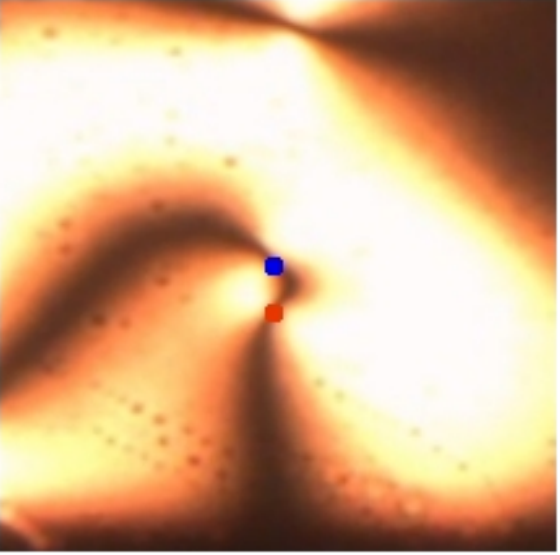}}
\hspace{0.35cm}
\subfloat[t=0 s]{\label{0}\includegraphics[scale=0.45]{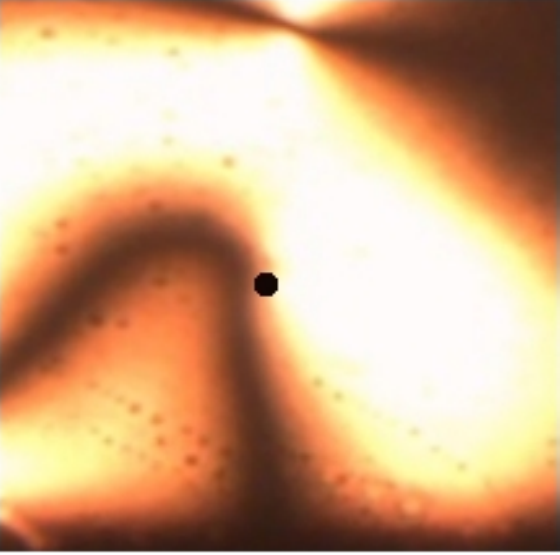}}
\caption{Typical photographic sequence showing a stringlike defect-antidefect pair annihilation in the KL/DeOH/H$_2$O mixture.}
\label{photoseq}
\end{figure}

As pointed before, individual defects of opposite sign attract each other and approach for an eventual annihilation. Figure \ref{photoseq} shows a typical texture sequence of approximation and annihilation of stringlike defects used for the determination of exponent $\alpha$ in equation (\ref{eq4}). Sequences of the same type were made for 49 initial configurations. Note that, after the annihilation, no defect remains in the region, showing that, indeed, the process occurs between a defect and an antidefect, i.e., $s_T = \sum s = -1/2 + 1/2 = 0$, where $s_T = 0$ represents a state without defects, exemplifying the conservation of the total topological charge $s_T$ \cite{lavrentovich}.

\begin{figure}[h]
\center
\label{dxt}\includegraphics[scale=1.020]{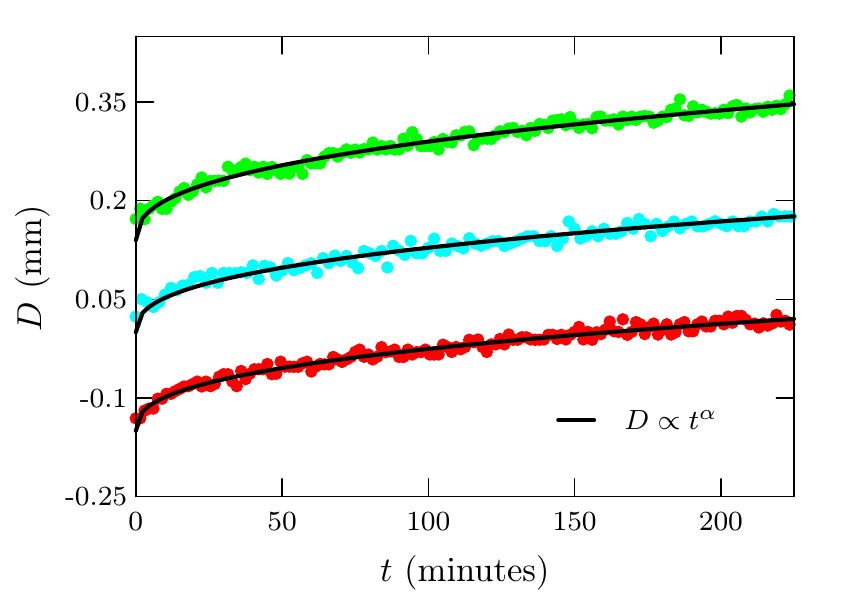}
\qquad
\hspace{-20pt}\label{mean}\includegraphics[scale=1.005]{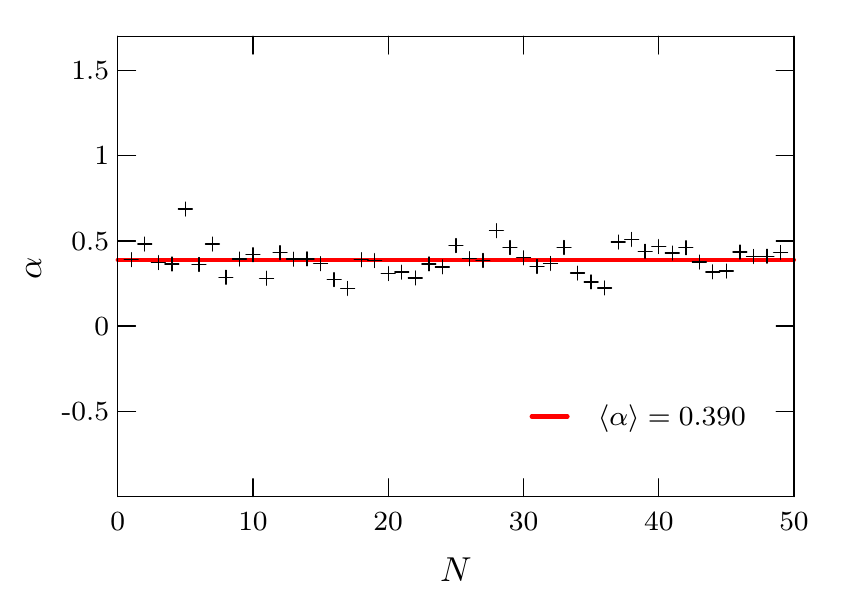}
\caption{(a) Three typical graphs, shifted vertically, showing the distance $D$ between two isolated defects \textit{versus} the time $t$ remaining for the annihilation. (b) Points represent exponents obtained from the annihilation dynamics of stringlike defects in the ternary mixture KL/DeOH/H$_2$O, and the continuous line corresponds to the average value $\alpha = 0.390$.}
\end{figure}

Representative results of the scaling law are given in figure \ref{dxt}, where the separation $D(t)$ between the defect and the antidefect in a pair is plotted as a function of the time $t$ remaining for the annihilation. When compared with $D \propto t^{\alpha}$, we obtained for the three annihilations in figure \ref{dxt} the mean value $\alpha = 0.390$. For these cases, the initial time $t$ was approximately 24000 seconds, the initial distance around 0.24 mm, and the time interval between the photos was 120 seconds. For distances lower than a few tenths of milimeters, the data became unreliable, being disregarded in the analysis. In that stage, the defect and the antidefect are annihilated, characterizing the end of the process (figure \ref{0}). Figure \ref{mean} shows the exponents obtained for our 49 analysis, leading to $\alpha = 0.390$ with standard deviation $\sigma = 0.085$.

Before comparing our results with others in the literature, we remark some general differences and similarities not previously cited between thermotropic and lyotropic liquid crystals. They are: lyotropic liquid crystals are more viscous than thermotropic ones \cite{jan, sampaio}; the order of magnitude of the birefringence for lyotropics is 10$^{-3}$ \cite{fneto} and for thermotropics it is 10$^{-1}$ \cite{degennes}; in phase transitions, thermotropics are sensitive to changes in temperature and pressure and exhibit a quite fast phase transition, whereas lyotropics are sensitive to changes in temperature, the relative concentration of the components of the mixture and pressure, with a phase transition more slower than in thermotropics \cite{fneto}.

In order to contextualize our findings, it is worthwhile to point out some previous results. In 1990, Shiwaku and co-authors \cite{shiwaku}, using thin films of main-chain-type thermotropic liquid-crystalline copolyester, found an exponent 0.35 for the annihilation between stringlike defects. In 1991, Pargellis and co-authors \cite{pargellis}, employing the thermotropic liquid crystal K15, found $\alpha = 0.5$; Chuang and co-authors \cite{turok} obtained $\alpha = 0.50 \pm 0.03$ by using the K15 liquid crystal. The exponent 0.5 was also found by Pargellis and co-authors \cite{finn} in 1992 for the annihilation between pointlike defects in a smectic thermotropic liquid crystal, but in the late stage, the exponent deviates towards a linear dependence with time. In 1993, Chuang and co-authors \cite{cypt} also obtained the exponent 0.50 $\pm$ 0.03 for the defect loop collapse in a thermotropic (5CB) liquid crystal. In 1994, Ding and Thomas \cite{ding} also obtained $D(t) \propto t^{0.5}$ for the annihilation of pointlike defects in a thermotropic liquid crystal polyester. In 1996, Pargellis and co-authors \cite{mendez}, now employing a thermotropic nematic known as E7, found that the distance between a pointlike defect pair in annihilation decreases as $t^{0.505 \pm 0.006}$. In 1997, Minoura and co-authors \cite{minoura} observed in a thermotropic 5CB that, in the late stage, the annihilation between pointlike defects goes with $t^{0.5}$. In 1998 \cite{minoura2}, they verified that this last result is independent of the alignment in the nematic cell; Wang and co-authors \cite{wang} found, for the shrinkage and annihilation of defects loop in a thermotropic liquid crystal polyester, that the radius $R$ of the defect is given by $R^2(t) \propto t$. In 2005, Dierking and co-authors \cite{marshall}, using a thermotropic liquid crystal, found that the distance between umbilical defects in annihilation decrease with $t^{0.49 \pm 0.01}$. In summary, except in the work of Shiwaku \cite{shiwaku}, an exponent around 0.5 seems to appear in most studies, and all of which were made in thermotropic or thermotropic polyesters liquid crystals. 

The annihilation between defects and the respective exponent associated has also been investigated through computer simulations. As mentioned before, in 1991, Pargellis and co-authors \cite{pargellis} call attention for a scaling solution for the separation between a pointlike defect-antidefect pair as $D \propto t^{1/2}$, using the nematodynamic equations \cite{degennes} and a model of forces in equilibrium $\rmd D/\rmd t \propto -1/D$. In 1992, Blundell and Bray \cite{blundell} investigated the effect of the modified symmetry in nematic liquid crystals on the ordering dynamics employing a model defined by $H = - \sum_{\langle i, j \rangle} (\phi_i \cdot \phi_j)^2$, where $\phi$ represents the director, and found, for three dimensions, that the characteristic length goes as $l(t) \sim t^{0.44 \pm 0.01}$ and the exponent decreases at later times. In 1993, Yurke and co-authors \cite{yurke} presented evidences that the coalescence of a point defect pair goes as $D^2 \ln D \sim t$ when the defect's size dependence is taken into account in its mobility. In 1995, Zapotocky and co-authors \cite{zapotocky} reported results from a cell-dynamical scheme simulation of phase ordering in two-dimensional samples of nematic liquid crystals. They found that the distance between stringlike defects decays as a power law with the exponent $\alpha = 0.375 \pm 0.007$ and this exponent seems to approach 0.5 when the average separation between defects becomes much larger than the cores. In another work in 1995, Zapotocky \cite{zapotocky2} also investigated the role played by topological textures during the phase ordering of an O(3) vector model in $d=2$ and found that the average separation $D(t)$ between topological objects goes as $t^{0.32 \pm 0.01}$. In 1999, Rojas and Rutenberg \cite{rojas} used simulations through the standard cell dynamical system (CDS) with dissipative dynamics and they found that the asymptotic law of the characteristic length scale is $L(t) \propto (t/ \ln [t/t_0])^{1/2}$ - result also obtained by Dutta and Roy \cite{dutta} in 2005 - and that the length scales can be well fitted employing an effective exponent of 0.40 $\pm$ 0.01. In 2001, Denniston and co-authors \cite{denniston}, using a lattice Boltzmann scheme, found that the effective exponent for the defect separation was around $\alpha = 0.47 \pm 0.03$. In 2006, Svetec and co-authors \cite{svetec} studied the semi-microscopic lattice-type model and Brownian molecular dynamics in an infinite cylinder of radius $R$ and were able to fit the annihilation regime between a pointlike defect pair with a single exponent, $D \propto t^{\alpha}$, with $\alpha = 0.4 \pm 0.1$. Finally, in 2010, Oliveira and co-authors \cite{oliveira} simulated the coarsening dynamics of defects in a two-dimensional nematic in the presence of an applied electric field founding that the characteristic length scales evolves roughly as $L^2 \propto t$, but change rapidly when the electric field is switched on.

Considering the standard deviation, our result indicates that the numerical simulations of Yurke and co-authors \cite{yurke}, Zapotocky and co-authors \cite{zapotocky}, and Svetec and co-authors \cite{svetec}, in addition to the experimental work of Shiwaku and co-authors \cite{shiwaku}, are in agreement with the results exhibited here. 

An important assumption made in deriving a relation for $D$ is that the evolution of average separation of the various defects formed after a phase transition is determined only by the forces acting in the isolated defect pair. The simulations made by Zapotocky confirmed this assumption. In our experiments, we inspected all the possible annihilations between isolated defect pairs separately. The good agreement of our result with the one found by Zapotocky \textit{et al.} \cite{zapotocky} shows consistency between this computer simulations and our experimental work. Moreover, most of the results reported here for the annihilation exponent $\alpha$ (figure \ref{mean}) were obtained under conditions when the pair separation $D$ was much smaller than the glass capillary thick.

We would like also to point out that the graphs in the figure \ref{dxt} and the ones for the other 46 pairs analyzed do not show a perceptible increase of the exponent $\alpha$, with increasing $D$. This feature is difficult to be analyzed experimentally due to the number of defects that show up at the sample.

\section{Conclusions}\label{conc}

In this work, we presented experimental results for the annihilation of isolated stringlike defects in a lyotropic liquid crystal with the following concentrations in weight percent: KL $\rightarrow$ 27.49\%, DeOH $\rightarrow$ 6.24\% and H$_2$O $\rightarrow$ 66.27\%. When compared with the power law $D \propto t^{\alpha}$, we obtained a scaling law with $\alpha = 0.390$ and standard deviation $\sigma = 0.085$, where $D$ is the distance between the annihilating defects, and $t$ the time remaining for the annihilation. Furthermore, it was observed only stringlike ($s= \pm 1/2$) defects in this analysis. Our results are in very good agreement to the ones obtained by Shiwaku and co-authors \cite{shiwaku}, Yurke and co-authors \cite{yurke}, Zapotocky and co-authors \cite{zapotocky} and Svetec and co-authors \cite{svetec}. In comparison to thermotropic liquid crystals, the lyotropic used here offers a good experimental condition to work with the defects, because the time of annihilation is very long and the attraction interaction between them seems to be weak. At least when based on the exponents $\alpha$, our investigations indicate that the annihilation dynamics in 2D of defects with strength $s = \pm 1/2$ is equivalent to those of type $s = \pm 1$ studied in \cite{oliveira} and \cite{svetec} through computer simulations. In addition, taking the error of experimental measurements into account, universality of $\alpha$ exponents in lyotropic and thermotropic can not be ruled out, although the average values of these exponents are different. We observed in our experiments that, after the annihilation, no defect was found in the region or, in other words, the sum over the defect's strength $s$ is zero. This conservation law illustrates a very general aspect of topological defects that occurs from early Universe to condensed matter. As far as we know, this is the first time that the stringlike defects dynamics is investigated using a lyotropic liquid crystal.

\ack
The authors thank B. F. de Oliveira for the useful discussions, comments and figure \ref{sketch}, and G. S. Dias for figure \ref{thermo}. This work was partially supported by Brazilian agencies (National Institutes of Science and Technology of Complex Fluids (INCT-FCx/CNPq) and Complex Systems (INCT-SC/CNPq), Funda\c{c}\~ao Arauc\'aria, CNPq and CAPES).


\section*{References}

\begin{thebibliography}{99}
\bibitem{degennes} de Gennes P G and Prost J 1995 \textit{The Physics of Liquid Crystals} 2nd ed (Clarendon, Oxford)
\bibitem{chandra} Chandrasekhar S 1980 \textit{Liquid Crystals} (Cambridge University Press, Cambridge)
\bibitem{lavrentovich} Kleman M and Lavrentovich O D 2003 \textit{Soft Matter Physics: An Introduction} (Springer, New York)
\bibitem{fneto} Figueiredo Neto A M and Salinas S R A 2005 \textit{The physics of lyotropic liquid crystals: phase transitions and structural properties} (Oxford University Press, Oxford, UK)
\bibitem{march} March N and Tosi M 1984 \textit{Polymers, Liquid Crystals and Low Dimensions Solids} (Plenum Press, New York)
\bibitem{martin} Martin J D, Keary C L, Thornton T A, Novotnak M P, Knutson J W and Folmer J C W 2006 \textit{Nature Materials} \textbf{5} 271
\bibitem{dierking} Dierking I 2003 \textit{Textures of Liquid Crystals} (Wiley-VCH, Weinheim)
\bibitem{chuang} Chuang I, Durrer R, Turok N and Yurke B 1991 \textit{Science} \textbf{251} 1336
\bibitem{turok} Chuang I, Turok N and Yurke B 1991 \textit{Phys. Rev. Lett.} \textbf{66} 2472
\bibitem{pargellis} Pargellis A, Turok N and Yurke B 1991 \textit{Phys. Rev. Lett.} \textbf{67} 1570
\bibitem{finn} Pargellis A N, Finn P, Goodby J W, Panizza P, Yurke B and Cladis P E 1992 \textit{Phys. Rev. A} \textbf{46} 7765
\bibitem{yurke} Yurke B, Pargellis A N, Kovacs T and Huse D A 1993 \textit{Phys. Rev. E} \textbf{47} 1525
\bibitem{toyoki} Toyoki H 1993 \textit{Phys. Rev. E} \textbf{47} 2558
\bibitem{zapotocky} Zapotocky M, Goldbart P M and Goldenfeld N 1995 \textit{Phys. Rev. E} \textbf{51} 1216
\bibitem{marshall} Dierking I, Marshall O, Wright J and Bulleid N 2005 \textit{Phys. Rev. E} \textbf{71} 061709
\bibitem{dutta} Dutta S and Roy S K 2005 \textit{Phys. Rev. E} \textbf{71} 026119
\bibitem{oliveira} de Oliveira B F, Avelino P P, Moraes F and Oliveira J C R E 2010 \textit{Phys. Rev. E} \textbf{82} 041707
\bibitem{kibble} Kibble T W B 1976 \textit{J. Phys. A - Math. and Gen.} {\bf 9} 1378
\bibitem{mermin} Mermin N D 1979 \textit{Rev. Mod. Phys.} {\bf 51} 591
\bibitem{charlier} Charlier J C, Ebbesen T W and Lambin P 1996 \textit{Phys. Rev. B} {\bf 53} 11108
\bibitem{vakaryuk} Vakaryuk V 2011 \textit{Phys. Rev. B} \textbf{84} 214524
\bibitem{figueiras} Figueiras C and de Oliveira B F 2011 \textit{Annalen Der Physik} \textbf{523} 898
\bibitem{petit} Petit-Garrido N, Trivedi N P, Ignes-Mullol J, Claret J, Lapointe C, Sagues F and Smalyukh I I 2011 \textit{Phys. Rev. Lett.} \textbf{107} 177801
\bibitem{abu} Abu-Libdeh N and Venus D 2011 \textit{Phys. Rev. B} \textbf{84} 094428
\bibitem{carvalho} Carvalho J, Furtado C and Moraes F 2011 \textit{Phys. Rev. A} \textbf{84} 032109
\bibitem{shiwaku} Shiwaku T, Nakai A, Hasegawa H and Hashimoto T 1990 \textit{Macromolecules} \textbf{23} 1590
\bibitem{blundell} Blundell R E and Bray A J 1992 \textit{Phys. Rev. A} \textbf{46} R6154
\bibitem{cypt} Chuang I, Yurke B, Pargellis A N and Turok N 1993 \textit{Phys. Rev. E} \textbf{47} 3343
\bibitem{ding} Ding D K and Thomas E L 1994 \textit{Mol. Cryst. Liq. Cryst.} \textbf{241} 103
\bibitem{zapotocky2} Zapotocky M and Zakrzewski W 1995 \textit{Phys. Rev. E} \textbf{51} R5189
\bibitem{mendez} Pargellis A N, Mendez J, Srinivasarao M and Yurke B 1996 \textit{Phys. Rev. E} \textbf{53} R25
\bibitem{minoura} Minoura K, Kimura Y, Ito K and Hayakawa R 1997 \textit{Mol. Cryst. Liq. Cryst.} \textbf{302} 345
\bibitem{minoura2} Minoura K, Kimura Y, Ito K and Hayakawa R 1998 \textit{Phys. Rev. E} \textbf{58} 643
\bibitem{wang} Wang W, Shiwaku T and Hashimoto T 1998 \textit{J. Chem. Phys.} \textbf{108} 1618
\bibitem{rojas} Rojas F and Rutenberg A D 1999 \textit{Phys. Rev. E} \textbf{60} 212
\bibitem{denniston} Denniston C, Orlandini E and Yeomans J M 2001 \textit{Phys. Rev. E} \textbf{64} 021701
\bibitem{svetec} Svetec M, Kralj S, Brada\v{c} Z and \v{Z}umer S 2006 \textit{Eur. Phys. J. E} \textbf{20} 71
\bibitem{digal} Digal S, Ray R and Srivastava A M 1999 \textit{Phys. Rev. Lett.} \textbf{83} 5030
\bibitem{mukai} Mukai H, Fernandes P R G, de Oliveira B F and Dias G S 2007 \textit{Phys. Rev. E} \textbf{75} 061704
\bibitem{pawel} Pieranski P 2009 On a Few Universal Aspects of Liquid Crystals \textit{P. G. de Gennes` Impact on Science} (Solid State and Liquid Crystals vol 1) ed J Bok, J Prost and F Brochard-Wyart (World Scientific) p 131
\bibitem{dhara} Dhara S, Arun Kumar T, Ishikawa K and Takezoe H 2009 \textit{J. Phys.: Condens. Matter} \textbf{21} 505103
\bibitem{blanc} Blanc C, Sven\v{s}ek D, \v{Z}umer S and Nobili M 2005 \textit{Phys. Rev. Lett.} \textbf{95} 097802
\bibitem{nagaya} Nagaya T, Hotta H, Orihara H and Ishibashi Y 1991 \textit{J. Phys. Soc. Jpn.} \textbf{60} 1572
\bibitem{svensek} Sven\v{s}ek D and \v{Z}umer S 2002 \textit{Phys. Rev. E} \textbf{66} 021712
\bibitem{gsdias} Dias G S, Fernandes P R G and Mukai H 2008 Estudos de Defeitos Topol\'ogicos em Sistemas L\'iquido Cristalinos In: 60a. Annual Meeting of the SBPC \textit{Proc. of the scientific meeting} Sao Paulo: SBPC/UNICAMP-Campinas-BR, available at: $<$http://www.sbpcnet.org.br/livro/60ra/resumos/resumos/R4500-1.html$>$. Accessed: February 9, 2013
\bibitem{renato} Guimar\~aes R R 2012 M.Sc. thesis Departamento de F\'isica Universidade Estadual de Maring\'a -  Brazil
\bibitem{barbero} Barbero G and Evangelista L R 2000 \textit{An Elementary Course on the Continuum Theory for Nematic Liquid Crystals} (World Scientific, Singapore)
\bibitem{kleman} Lavrentovich O D and Kleman M 2001 \textit{Chirality in Liquid Crystals} (Springer, New York)
\bibitem{kroin} Kroin T, Palangana A J and Figueiredo Neto A M 1989 \textit{Phys. Rev. A} \textbf{39} 5373
\bibitem{evangelista} Barbero G and Evangelista L R 2006 \textit{Adsorption phenomena and anchoring energy in nematic liquid crystals} (Taylor \& Francis, London)
\bibitem{jan} Jadzin J and Czechowski G 2001 \textit{J. Phys.: Condens. Matter} \textbf{13} L261
\bibitem{sampaio} Sampaio A R, Fernandes P R G, Simoes M and Palangana A J 2001 \textit{Mol. Cryst. Liq. Cryst.} \textbf{359} 589
\end{thebibliography}
\end{document}